\documentclass[twocolumn,showpacs,preprintnumbers,amsmath,amssymb,superscriptaddress]{revtex4-1}

\usepackage[colorlinks,pdfstartview=FitH]{hyperref}
\hypersetup{linkcolor=blue,citecolor=blue,filecolor=black,urlcolor=blue}

\usepackage{amsmath,amssymb,bm}
\usepackage{graphicx}
\usepackage{amsfonts}
\usepackage{color}

\begin{document}

%\begin{CJK*}{GBK}{song}

\baselineskip=15pt \parskip=5pt

\vspace*{3em}

\title{Explanation of the 511 keV line: Cascade annihilating dark matter with the $^8$Be anomaly}

\author{Lian-Bao Jia }
\email{jialb@mail.nankai.edu.cn}
\affiliation{School of Science, Southwest University of Science and Technology, Mianyang
621010, China}

\begin{abstract}

A possible dark matter (DM) explanation about the long-standing issue of the Galactic 511 keV line is explored in this paper. For DM cascade annihilations of concern, a DM pair $\pi_d^{+} \pi_d^{-}$ annihilates into unstable $\pi_d^{0} \pi_d^{0}$, and $\pi_d^{0}$ decays into $e^+ e^-$ with new interactions suggested by the $^8$Be anomaly. Considering the constraints from the effective neutrino number $N_{\mathrm{eff}}$ and the 511 keV gamma-ray emission, a range of DM is obtained, $11.6 \lesssim  m_{\pi_d^{\pm}} \lesssim  15$ MeV. The typical DM annihilation cross section today is about 3.3 $\times$ $10^{-29}$ cm$^3$ s$^{-1}$, which can give an explanation about the 511 keV line. The MeV scale DM may be searched by the DM-electron scattering, and the upper limit set by the CMB s-wave annihilation is derived in DM direct detections.

\end{abstract}

\maketitle

%\end{CJK*}

\section{Introduction}

It has been many decades since the observation of the Galactic 511 keV gamma-ray line in 1970s \cite{Leventhal:1978} (see Ref. \cite{Prantzos:2010wi} for a review), while the source of it is still unclear. It was demonstrated by INTEGRAL \cite{Winkler:2003nn} in 2003 that the 511 keV line originates from the electron-positron annihilations in positronium states. The emission was found to be concentrated towards the Galactic bulge region \cite{Knodlseder:2005yq}, and few point-like sources have been detected by INTEGRAL \cite{Ubertini:2014nla,Siegert:2015knp}. A possible explanation of the 511 keV line is from dark matter (DM) annihilations \cite{Boehm:2003bt,Fayet:2004bw,Boehm:2004gt,Beacom:2004pe,Ascasibar:2005rw} or decays \cite{Hooper:2004qf,Picciotto:2004rp}, or new atoms binded \cite{Cudell:2014eta}. Analysis results of Refs. \cite{Ascasibar:2005rw,Vincent:2012an} show that the INTEGRAL signals are more peaked towards the center than the DM density profiles, and the decaying DM origin of $e^+$ is disfavored. For DM annihilations, the interesting scenario of a pair of DM particles annihilating into a pair of $e^+ e^-$ with the DM mass $\lesssim$ 7.5 MeV and the annihilation cross section $\sim$ $10^{-30}$ cm$^3/$s is allowed by the continuum gamma-ray emission \cite{Beacom:2005qv,Sizun:2006uh}, while this scenario seems in tension with the present cosmological data \cite{Wilkinson:2016gsy}. In addition, the signal is difficult to be explained by the known astrophysical sources \cite{Prantzos:2010wi,Wilkinson:2016gsy,Davoudiasl:2009ud}. Thus, the main source of the 511 keV emission line is unclear yet.

As the transition mechanism between DM and the standard model (SM) particles is unknown, exotic DM sources can still be a reasonable explanation about the 511 keV line, and possible signatures of new interactions may bridge the hidden sector and the SM sector. Recently, an apparent anomaly with 6.8$\sigma$ discrepancy was observed in $^8$Be transitions \cite{Krasznahorkay:2015iga}, and a new vector boson $X$ was suggested, with the mass $m_X \simeq$ 17 MeV and the transition form $X \to e^+ e^-$. For possible couplings of $X$ with quarks and leptons, the vector form was studied in Refs. \cite{Feng:2016jff,Feng:2016ysn}, and the axial vector form was analyzed in Ref. \cite{Kozaczuk:2016nma} (for more discussions, see e.g., Ref. \cite{Gu:2016ege}). For the possible $X$ portal DM investigated in Refs. \cite{Jia:2016uxs,Kitahara:2016zyb,Chen:2016tdz,Seto:2016pks}, the DM in ten MeV scale is allowed by the constraints. Apparently, the direct annihilation mode of a DM pair $\to {X} \to$ $e^+ e^-$ is faint for the 511 keV line.

The $X$ portal DM could potentially be the source of the 511 keV line, while the injection of the low energy $e^+$ becomes a crucial question. This is of our concern in this paper. In the case of a pair of DM particles first annihilating into unstable particles, then converting into SM particles, the required low energy $e^+$ may be obtained in cascade annihilations (for more about this type annihilations, see e.g., Refs. \cite{Martin:2014sxa,Abdullah:2014lla,Elor:2015tva}). Here we consider a scenario of three dark pions are in a triplet in the hidden sector, with two dark charged pions being DM candidates and the unstable neutral pion transiting into SM particles via the $X$ boson. The main annihilation channel is a pair of dark charged pions  $\to $ a pair of neutral dark pions which converting into SM $e^+ e^-$ pairs during DM freeze-out, and this mode is phase space suppressed today. In this scenario, the required low energy $e^+$ can be obtained for the 511 keV line. In addition, the constraint from the cosmic microwave background (CMB) observation \cite{Ade:2015xua,Slatyer:2015jla} will be considered in DM annihilations.

For DM with the mass $\sim$ 10 MeV, the dark sector energy release to the big bang nucleosynthesis (BBN) and the effective number of relativistic neutrino $N_{\mathrm{eff}}$ at the recombination needs to be considered, and the parameter space will be derived with these constraints. For the ten-MeV scale DM, the DM-electron scattering can be employed in the search of DM \cite{Bernabei:2007gr,Dedes:2009bk,Kopp:2009et}, and the direct detection of DM will be discussed.

The rest of this paper is organized as follows. The interactions in the new sector will be presented, and the annihilations of DM will be discussed in the next. Then, the numerical analysis with constraints are elaborated. The last part is the conclusion.

\section{Interactions in the new sector}

Now we briefly introduce the model. We suppose a SU(N)$_d$$\times U(1)_X$ gauge symmetry in the hidden sector, with the SU(N)$_d$ interactions being similar to the strong interactions in SM (see e.g., Ref. \cite{Ryttov:2008xe} for more). Two SM-singlet light dark quarks $u_d$ and $d_d$ are assumed to be coupled under $U(1)_X$. With couplings to a dark Higgs field, the mass of the vector boson $X$ is obtained, and the light dark quarks $u_d$ and $d_d$ are assumed with equal masses. At the energy below the dark QCD-like scale $\Lambda_N$, the dark chiral symmetry of the light dark quarks is broken and three dark pions are generated, with the two dark charged pions $\pi_d^{\pm}$ being DM candidates and the neutral dark pion $\pi_d^{0}$ being unstable. Here the value $\Lambda_N  \sim  f_{\pi_d}$ is taken in the dark sector, and $f_{\pi_d}$ is the dark pion decay constant. For more about a similar case, see the dark pions considered in Ref. \cite{Kopp:2016yji}. In addition, SM particles are singlets under SU(N)$_d$. The SM fermions suggested by the $^8$Be anomaly \cite{Krasznahorkay:2015iga} couple to the $X$ boson, and the effective interaction mediated by $X$ is taken as $\mathcal{L}_I^i$ =  $- X_\mu J^\mu_{\mathrm{SM}}$, where the SM fermion current $J^\mu_{\mathrm{SM}}$ is
\begin{eqnarray}
J^\mu_{\mathrm{SM}} = \sum_f  \bar{f} (g^V_f \gamma^\mu + g^A_f \gamma^\mu \gamma^5)  f    ~.  \label{X-SM-C}
\end{eqnarray}
For interactions mediated by $X$, both vector and axial couplings are included for SM fermions. For the excited states of Beryllium, the anomaly was observed in the transition of isoscalar $^8\mathrm{Be}^\ast$, while the isovector ${^8\mathrm{Be}^\ast}^\prime$ was not. This can be naturally explained by the primary axial couplings of $X$ to quarks \cite{Feng:2016jff,Feng:2016ysn,Kozaczuk:2016nma}, and the couplings suggested by the $^8$Be anomaly are \cite{Kozaczuk:2016nma}
\begin{eqnarray}
&10^{-5} \lesssim \mathrm{max}(|g^A_u|,|g^A_d|) \lesssim 10^{-4}   ~,  \\
&2 \times 10^{-4}  \lesssim \sqrt{(g^A_e)^2 + (g^V_e)^2 }/e  \lesssim 2 \times 10^{-3}  \nonumber \\
&\quad \mathrm{and} \quad      |g^V_e g^A_e| \lesssim 1 \times 10^{-8}    ~.  \label{x-e-coupling}
\end{eqnarray}

Consider the leptonic decay of $\pi_d^{0}$ mediated by $X$. In the process $\pi_d^{0}$ $\to e^+ e^-$, the corresponding matrix element is taken as
\begin{eqnarray}
\langle e^+ e^- | H_X | \pi_d^{0} \rangle  = - \frac{\lambda^A_d}{m_X^2} \langle e^+ e^- | J^\mu_{e^+ e^-} | 0 \rangle \langle 0 | A^d_\mu| \pi_d^{0} \rangle  ~,
\end{eqnarray}
where $\lambda^A_d$ is the axial vector coupling in the $X$$-$$\pi_d^{0}$ interaction. The annihilating matrix element of $\pi_d^{0}$ is written in the form
\begin{eqnarray}
\langle 0 | A^d_\mu (x) | \pi_d^{0} \rangle  = i f_{\pi_d} k_\mu e^{-i k \cdot x } ~,
\end{eqnarray}
where $k_\mu$ is the momentum of $\pi_d^{0}$. The matrix element $\langle e^+ e^- | J^\mu_{e^+ e^-} | 0 \rangle$ can be obtained via Eq. (\ref{X-SM-C}). Provided that the channel $\pi_d^{0}$ $\to X^\ast \to e^+ e^-$ is dominant in $\pi_d^{0}$'s decay, the decay width of $\pi_d^{0}$ is
\begin{eqnarray}
\Gamma(\pi_d^{0}) \approx \frac{(\lambda^A_d )^2  (g^A_e )^2 f_{\pi_d}^2}{2 \pi m_{X}^4}  m_{\pi_d^{0}} m_e^2  (1 - \frac{4 m_e^2}{m_{\pi_d^{0}}^2})^{1/2}   ~.
\end{eqnarray}

The effective interaction between the dark scalar field $\Phi$ and dark pions are taken as
\begin{eqnarray}
\mathcal{L}_{\Phi}^i &=& - \frac{1}{2} \lambda \Phi^2 \pi_d^{+} \pi_d^{-} -  \frac{1}{4} \lambda_0 \Phi^2 \pi_d^0 \pi_d^0 \nonumber \\
&& - \mu \Phi \pi_d^{+} \pi_d^{-}  - \frac{1}{2} \mu_0 \Phi \pi_d^0 \pi_d^0 ~.
\end{eqnarray}
Here the values $\lambda_0 = \lambda$ and $\mu_0 = \mu$ are taken for simplicity. The mass difference $\Delta$ between $\pi_d^{\pm}$ and $\pi_d^{0}$ can be tiny for a very weak interaction in forms of the dark charge \cite{Das:1967it}. The dark scalar could also have a weak coupling to SM particles, e.g., via a mixing with SM Higgs field \cite{Jia:2017kjw}, and a new particle $\phi$ (in the mass eigenstate) couples to SM particles. The effective interaction of $\phi$ with the SM fermion is parameterized as
\begin{eqnarray}
\mathcal{L}_{\phi}^i = - \theta \frac{m_f}{v} \bar{f} f \phi ~,
\end{eqnarray}
where the vacuum expectation value $v$ is $v \approx$ 246 GeV, and $\theta$ is a parameter with $|\theta| \ll 1 $.

Here the particles we focus on are relevant to the main interactions between DM and SM particles, and we should keep in mind that there may be more particles in the dark sector.

\section{Annihilations of DM}

We consider that the cascade annihilation process $\pi_d^{+} \pi_d^{-}$  $\to$ $\pi_d^{0} \pi_d^{0}$ with $\pi_d^{0}$ decaying into $e^+ e^-$ is dominant during DM freeze-out, which mainly contributes by the scattering of dark pions. The dark pion scattering amplitude \cite{Weinberg:1966kf} is $\mathcal{M}(\pi_d^{+} \pi_d^{-} \to \pi_d^{0} \pi_d^{0})$ = $(s - m_{\pi_d}^2) / f_{\pi_d}^2$, and the corresponding annihilation cross section is
\begin{eqnarray}
\sigma_0^{} v_r  = \frac{1}{2} \frac{ \beta_0 }{ 32 \pi (s - 2 m_{\pi_d^{\pm}}^2)}  \frac{(s -  m_{\pi_d^{\pm}}^2)^2}{f_{\pi_d}^4} ~,
\end{eqnarray}
where $v_r$ is the relative velocity between the two DM particles, and the factor $\frac{1}{2}$ is due to the required $\pi_d^{+} \pi_d^{-}$ pair in DM annihilations. $s$ is the total invariant mass squared, and the factor $\beta_0$ is $\beta_0 =$ $ \sqrt{1 - 4 m_{\pi_d^{0}}^2 / s}$. Here we consider that the $\sigma_0^{} v_r$ gives the main contribution to the annihilation cross section during DM freeze-out, and the value of $f_{\pi_d}$ can be obtained with the DM relic density inputted. The DM cascade annihilation is phase space suppressed today. In the nonrelativistic limit, one has $s$ = 4$m_{\pi_d^{\pm}}^2 + m_{\pi_d^{\pm}}^2 v_r^2 + \mathcal{O} (v_r^4)$, and
\begin{eqnarray}
\beta_0  \approx \sqrt{{v_r^2}/{4} + 2{\Delta}/{m_{\pi_d^{\pm}}}} ~ .
\end{eqnarray}
For the Galactic 511 keV line due to DM cascade annihilations, as the energy of the injected positron should be $\lesssim$ 7.5 MeV \cite{Beacom:2005qv,Sizun:2006uh}, this means that the DM mass should be $m_{\pi_d^{\pm}}$ ($\simeq$ $m_{\pi_d^{0}}$) $\lesssim$ 15 MeV.

We assume that the lifetime of $\pi_d^{0}$ is less than (or similar to) the age of the early universe when the temperature $T$ of the universe cooling to $m_{\pi_d^{0}}$. In this case, the number densities of dark pions are of their equilibrium values before DM freeze-out. At $T \sim$ $m_{\pi_d^{0}}$, the effective lifetime $\tau_{eff}$ of $\pi_d^{0}$ is
\begin{eqnarray}
\frac{1}{\tau_{eff}} \simeq \frac{m_{\pi_d^{0}}}{\langle  E_{\pi_d^{0}} \rangle} \Gamma(\pi_d^{0}) ~,
\end{eqnarray}
where the averaged energy of $\pi_d^{0}$ is $\langle  E_{\pi_d^{0}} \rangle \approx$ 3.25 $T$. Thus, we can derive a limit (see e.g., Ref. \cite{Jia:2017kjw} for more)
\begin{eqnarray}
(\lambda^A_d )^2  (g^A_e )^2 f_{\pi_d}^2 \gtrsim 1.775 \times  10^{-21} \sqrt{g_\ast} m_{\pi_d^{0}} (\mathrm{MeV}) ~ ,
\end{eqnarray}
with $f_{\pi_d}$ and $m_{\pi_d^{0}}$ in units of GeV and MeV, respectively. $g_\ast$ is the effective number of the relativistic degrees of freedom.

For the s-wave process $\pi_d^{+} \pi_d^{-}$ $\to {\phi} \to$ $e^+ e^-$, in the case of $m_{\pi_d^{\pm}} < m_{\phi} \ll m_h$, the annihilation cross section is
\begin{eqnarray}
\sigma_1^{} v_r  \approx  \frac{1}{2} \frac{ \beta_1 (s - 4 m_e^2)  m_e^2  }{ 8 \pi (s - 2 m_{\pi_d^{\pm}}^2) v^2 } \frac{\theta^2  \mu^2 }{  (s - m_{\phi}^2)^2 }  ~,
\end{eqnarray}
where the phase space factor $\beta_1$ is $\beta_1 =$ $ \sqrt{1 - 4 m_e^2 / s} $. For this s-wave annihilation as small as possible, we consider that the annihilation is away from the resonance, e.g.,
\begin{eqnarray}
|1 -   2 m_{\pi_d^{\pm}} / m_{\phi} | \gtrsim  0.15 ~. \label{cmb-up}
\end{eqnarray}
The CMB observation sets a stringent constraint on s-wave annihilations of DM in MeV scale, and this constraint will be considered in the following.

\section{Numerical analysis}

\subsection{DM mass with constraints of $N_{\mathrm{eff}}$}

For DM with the mass $m_{\pi_d^{\pm}}$ $\lesssim$ 15 MeV, the main annihilation product in SM sector is $e^+ e^-$. For thermal freeze-out DM, the value of the parameter $x_f = m_{\pi_d^{\pm}} / T_f$ is about 20, where $T_f$ is the freeze-out temperature of DM. This means that the electron-photon plasma in the early universe was heated by the energy injected from dark pions after the decoupling of the electron neutrinos (a typical decoupling temperature $T_d$ for neutrinos is $T_d \sim$ 2.3 MeV \cite{Enqvist:1991gx}). The abundance of light elements from BBN and the CMB power spectra at recombination time can be altered by the energy injection. These effects can be considered in terms of the effective number of the relativistic neutrinos $N_{\mathrm{eff}}$, and the expectation value is $N_{\mathrm{eff}}$ = 3.046 in the standard cosmological prediction \cite{Dolgov:2002wy,Mangano:2005cc}. Considering the constrains on $N_{\mathrm{eff}}$ with the Planck data \cite{Ade:2015xua}, a lower bound $ N_{\mathrm{eff}} \gtrsim $ 2.9 is adopted \cite{Jia:2016uxs}. Moreover, as the mass $m_X \gg T_d$ and the lifetime of $X$ $\ll$ 1 second, the contribution from $X$ boson is negligible.

\begin{figure}[htbp]
\includegraphics[width=0.37\textwidth]{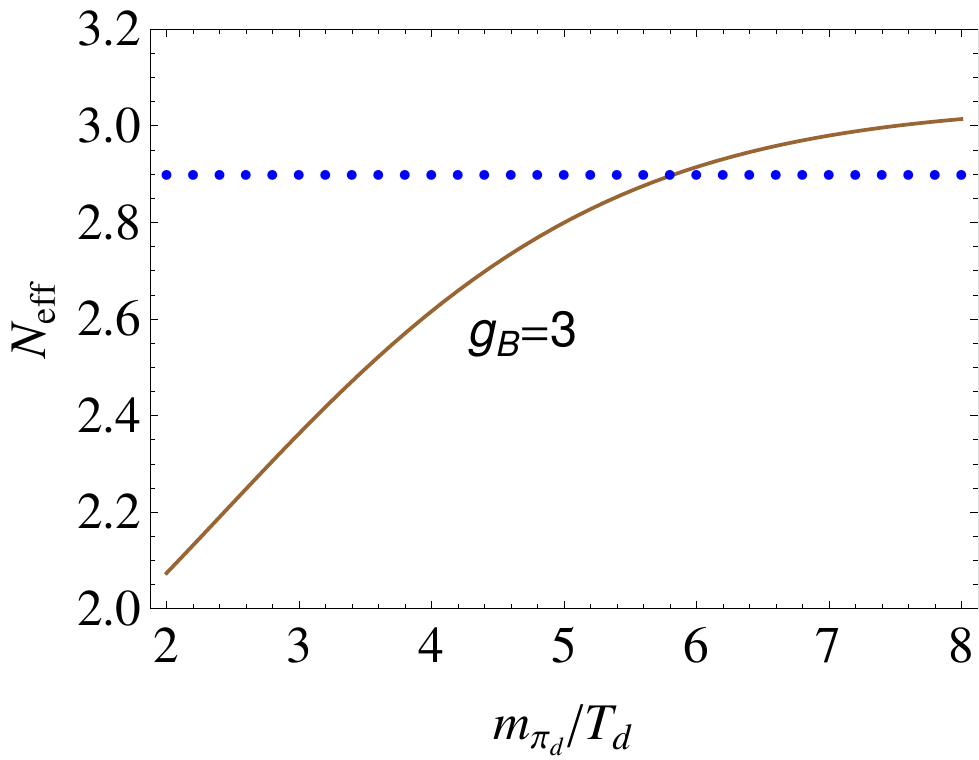} \vspace*{-1ex}
\caption{The effective number $N_{\mathrm{eff}}$ as a function of $m_{\pi_d} / T_d$. The solid curve is for dark pions with $g_B^{}$ = 3. The dotted curve is for the lower bound $N_{\mathrm{eff}}$ = 2.9. }\label{neff-d}
\end{figure}

The DM particles $\pi_d^{+} \pi_d^{-}$ freeze out later than the neutrino decoupling, and the $\pi_d^{0}$ particles keep in the thermal equilibrium with DM particles when the temperature cooling to $T_d$. For $\Delta / T_d \ll 1$, the number density of DM particles is nearly twice of that of $\pi_d^{0}$ particles at the time of the neutrino decoupling. The energy injected from dark pions heats the electron-photon plasma, and the effective number $N_{\mathrm{eff}}$ is written in forms of \cite{Ho:2012ug,Ho:2012br}
\begin{eqnarray}
N_{\mathrm{eff}}  = 3.046 \bigg[ \frac{I(0)}{I(T_d)} \bigg]^{\frac{4}{3}}  ~,
\end{eqnarray}
where the function $I(T_\gamma)$ is
\begin{eqnarray}
I(T_\gamma) &\simeq&  \frac{1}{T_\gamma^4}  (\rho_{e^+ e^-} + \rho_{\gamma} + \rho_{\pi_d} + p_{e^+ e^-} + p_{\gamma} + p_{\pi_d})     \nonumber  \\
&\simeq& \frac{11}{45} \pi^2 + \frac{g_B^{}}{2 \pi^2} \int_{y=0}^{\infty} \mathrm{d}y \frac{y^2}{e^{\xi} - 1} (\xi + \frac{y^2}{3 \xi})   ~ ,
\end{eqnarray}
with $\xi$ = $\sqrt{y^2 + (m_{\pi_d} / T_{\gamma})^2}$. Here $T_{\gamma}$ is the temperature of the photon, and the integration variable $y$ is equal to $p_{\pi_d} / T_{\gamma}$. For dark pions, the degree of freedom is $g_B^{}$ = 3. The result of the effective number $N_{\mathrm{eff}}$ as a function of $m_{\pi_d} / T_d$ is depicted in Fig. \ref{neff-d}. Considering the lower bound $N_{\mathrm{eff}} \gtrsim$ 2.9, we obtain the result $m_{\pi_d} / T_d \gtrsim$ 5.8 for dark pions. As the neutrino decoupling is not a sudden process \cite{Enqvist:1991gx,Dolgov:2002wy,Mangano:2005cc,Hannestad:2001iy}, a lower bound $T_d \gtrsim$ 2 MeV is adopted here. Now, taking $m_{\pi_d^{\pm}}$ as the value of $m_{\pi_d}$, the mass range of DM is
\begin{eqnarray}
11.6 \lesssim  m_{\pi_d^{\pm}} \lesssim  15  ~~  (\mathrm{MeV})   ~  .
\end{eqnarray}

In addition, for the alteration of the BBN and CMB being as small as possible, here we consider the case that masses of the massive mediators are larger than the dark pion mass, e.g., with masses not smaller than $m_X$. In this case, the Yukawa-type potentials cannot significantly affect the two-body wavefunction in DM nonrelativistic annihilations \cite{ArkaniHamed:2008qn}, and the corresponding Sommerfeld enhancements are negligible.

\subsection{Parameters for the 511 keV line}

\begin{figure}[htbp]
\includegraphics[width=0.37\textwidth]{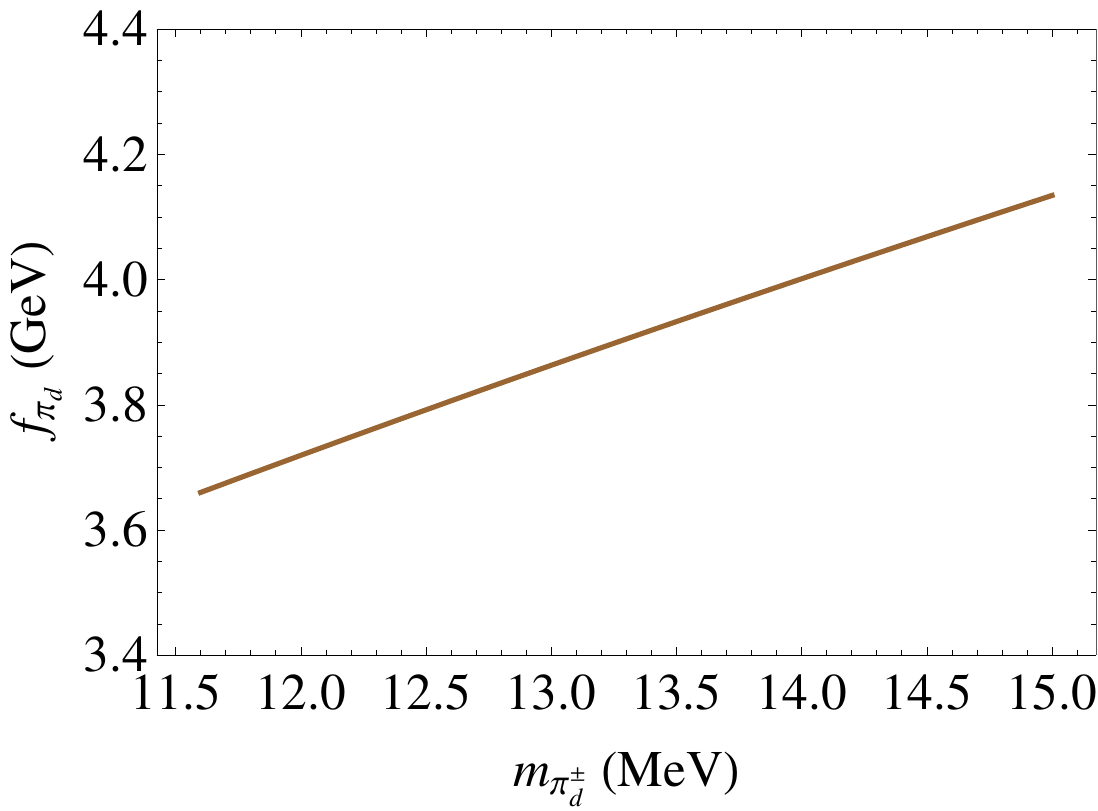} \vspace*{-1ex}
\caption{The value of $f_{\pi_d}$ as a function of $m_{\pi_d^{\pm}}$. }\label{dark-f}
\end{figure}

Consider the constraint from DM relic abundance first. The relic density of DM is $\Omega_{D} h^2$ = 0.1197$\pm$0.0042 \cite{Ade:2015xua}. Here the scattering of dark pions gives the main contribution to DM annihilations (i.e., $\sigma_0^{} v_r $) during DM freeze-out. The result of $f_{\pi_d}$ as a function of $m_{\pi_d^{\pm}}$ is shown in Fig. \ref{dark-f}. After the value of $f_{\pi_d}$ being derived, in the nonrelativistic limit, one has that the annihilation of DM is about
\begin{eqnarray}
2 \times 10^{-25} \times \beta_0 ~ \mathrm{cm}^3/s
\end{eqnarray}
for the DM mass range of concern.

Now, we turn to the constraint from CMB. The CMB constraint on the annihilation mode of a DM pair $\to$ $e^+ e^-$ was derived in Ref. \cite{Slatyer:2015jla}. For the DM of concern, the revised upper limit of the similar mode can be obtained via the DM mass multiplied by two and the corresponding upper limit doubled compared with that in Ref. \cite{Slatyer:2015jla}, i.e., the annihilation cross section of the process $\pi_d^{+} \pi_d^{-}$  $\to$ $\pi_d^{0} \pi_d^{0}$ should be $\lesssim$ (6$-$7)$\times 10^{-30}$ cm$^3$/s at the recombination. In the DM temperature $\sim$ 0 limit, $\Delta / m_{\pi_d^{\pm}} \lesssim 5 \times 10^{-10}$ is adopted. The dark charge coupling parameter $\alpha_d^{}$ (like $\alpha$ in SM) is related to the mass difference $\Delta$ \cite{Kopp:2016yji,Das:1967it}, and this means $\alpha_d^{} \lesssim$ $10^{-13}$ with the constraint of CMB. Thus, we neglect the contribution via dark charge couplings in DM annihilations and self-interactions.

In DM cascade annihilations, to obtain the Galactic 511 keV line, the required annihilation cross section is \cite{Boehm:2004gt}
\begin{eqnarray}
\langle \sigma_0^{} v_r \rangle \sim   4.32 \times  10^{-29}  \big [\frac{ m_{\pi_d^{\pm}}  (\mathrm{MeV})}{10 } \big ]^2 \mathrm{cm}^3 \mathrm{s}^{-1} ,
\end{eqnarray}
with $m_{\pi_d^{\pm}}$ in units of $\mathrm{MeV}$. The typical relative velocity near the Galactic bulge is $v_r \sim $ 100 km/s \cite{Jia:2017kjw}. Thus, the typical DM annihilation cross section is about 3.3 $\times$ $10^{-29} \mathrm{cm}^3 \mathrm{s}^{-1}$. The DM annihilation cross section is of the order required by the 511 keV line, and the DM of concern can give an explanation about the 511 keV line. Moreover, the DM annihilation cross section is tolerant by the constraint from dwarf satellite galaxies \cite{Siegert:2016ijv}.

\subsection{Direct detections of DM}

Due to the small recoil energy, the present DM-target nucleus scattering experiment is insensible for the search of MeV scale DM. Here, we turn to the DM-electron scattering. In the DM-electron scattering, the typical momentum transfer $q$ is of order $\alpha m_e$, with the recoil energy of a few eV. The form of DM-electron scattering given by Ref. \cite{Essig:2011nj} is adopted. Here, the DM-electron scattering cross section is
\begin{eqnarray}
\bar{\sigma}_e  &=&  \frac{\mu_{\pi_d^{\pm} e}^2}{16 \pi m_{\pi_d^{\pm}}^2 m_e^2}  \overline{|\mathcal{M}_{\pi_d^{\pm} e} (q)|^2} |_{q^2 = \alpha^2 m_e^2}   \times  |F_{\mathrm{DM}} (q)|^2   \nonumber  \\
&\approx& \frac{\mu_{\pi_d^{\pm} e}^2}{4 \pi m_{\pi_d^{\pm}}^2}  \frac{m_e^2}{v^2}   \frac{\theta^2  \mu^2 }{  m_{\phi}^4 }  ~,
\end{eqnarray}
where $\mu_{\pi_d^{\pm} e}$ is the $\pi_d^{\pm}$-electron reduced mass, and $F_{\mathrm{DM}} (q)$ $\simeq $1 is taken for $m_{\phi} \gg \alpha m_e $.

\begin{figure}[htbp]
\includegraphics[width=0.4\textwidth]{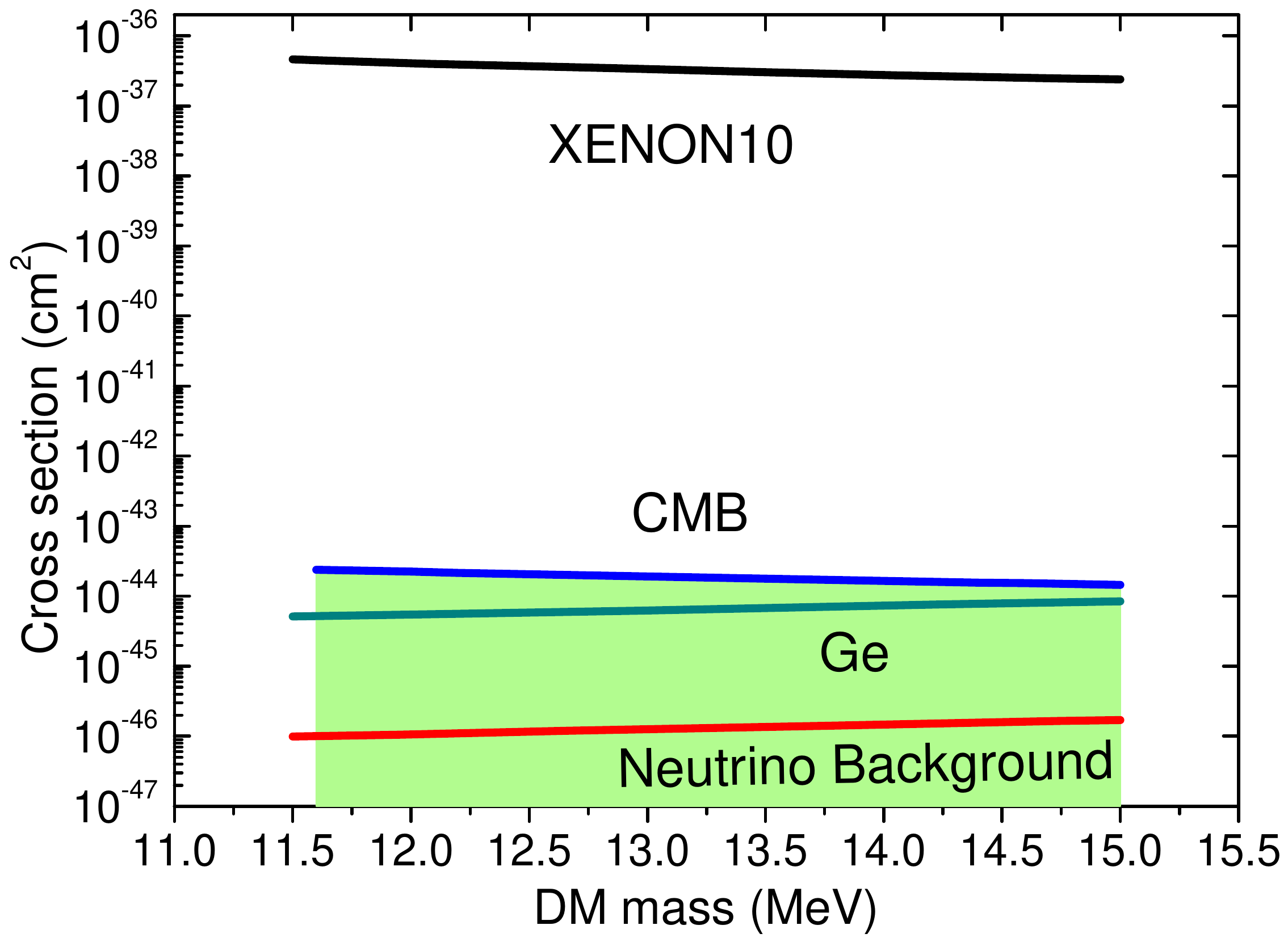} \vspace*{-1ex}
\caption{The DM-electron scattering cross section $\bar{\sigma}_e$ as a function of DM mass. The curves from top to bottom are the upper limit set by the XENON10 \cite{Essig:2012yx}, the s-wave annihilation constraint from CMB, the Ge detector scattering cross section in a 1 kg-year exposure \cite{Mei:2017etc}, and the lower detection bound set by the neutrino background \cite{Billard:2013qya}, respectively. The shaded region is the range of possible values of $\bar{\sigma}_e$, with the upper limit set by CMB (here $ \theta^2  \mu^2  / m_{\phi}^4$  $\lesssim$ 0.1 is taken).}\label{dark-e}
\end{figure}

Now, consider the CMB constraint on DM s-wave annihilations, and take it into consideration in DM direct detections. For the s-wave annihilation $\pi_d^{+} \pi_d^{-}$ $\to {\phi} \to$ $e^+ e^-$, the annihilation cross section should be $\lesssim  10^{-30}$ cm$^3$/s with the constraint from CMB. Thus, we have
\begin{eqnarray}
\frac{\theta^2  \mu^2 }{ (4 m_{\pi_d^{\pm}}^2 - m_{\phi}^2)^2 } \lesssim 1   ~,
\end{eqnarray}
with masses in units of GeV. Considering Eq. (\ref{cmb-up}), we have $ \theta^2  \mu^2  / m_{\phi}^4$  $\lesssim$ 0.1. The numerical result of $\bar{\sigma}_e$ is depicted in Fig. \ref{dark-e}. For the DM-electron scattering, it can be seen that, the s-wave annihilation constraint from CMB is much lower than the upper limit set by the XENON10 \cite{Essig:2012yx}. For the DM of concern, the small $\bar{\sigma}_e $ can be explored by e.g., the Ge detector at low temperature of a few K, and the scattering cross section can reach about (5$-$8)$\times 10^{-45}$ cm$^2$ in a 1 kg-year exposure \cite{Mei:2017etc}. The lower detection bound from the neutrino background is taken as that in Ref. \cite{Billard:2013qya}.

\section{Conclusion}

A DM interpretation of the Galactic 511 keV line has been demonstrated in this article, i.e., a DM pair $\pi_d^{+} \pi_d^{-}$ annihilates into unstable $\pi_d^{0} \pi_d^{0}$, and $\pi_d^{0}$ decays into $e^+ e^-$ with new interactions suggested by the $^8$Be anomaly. Considering the constraint from $N_{\mathrm{eff}}$ and the 511 keV gamma-ray emission, a mass range of DM is obtained, $11.6 \lesssim  m_{\pi_d^{\pm}} \lesssim  15$ MeV. For the DM annihilations today, the typical DM annihilation cross section is about 3.3 $\times$ $10^{-29} \mathrm{cm}^3 \mathrm{s}^{-1}$, which is of the order required by the 511 keV line. Thus, the DM of concern can give an explanation about the 511 keV line.

The MeV scale DM can be searched by the DM-electron scattering. In this paper, we consider the DM-electron scattering mainly mediated by $\phi$, and an upper limit of this scattering is set by the CMB s-wave annihilation (for the DM s-wave annihilation mediated by $\phi$), with $\bar{\sigma}_e \lesssim 10^{-44}$ cm$^2$. The Ge detector at low temperature of a few K can be employed to search for MeV scale DM, with the scattering cross section $\bar{\sigma}_e$ reaching about (5$-$8)$\times 10^{-45}$ cm$^2$ in a 1 kg-year exposure \cite{Mei:2017etc}. For the DM of concern, the lower detection bound from the neutrino background may be reached in about 50 kg-year exposure by the Ge detector. Moreover, if there is a vectorial interaction between scalar DM and $X$, as discussed in Ref. \cite{Jia:2016uxs}, the CMB constraint on DM-electron scattering may be relaxed. Briefly, the DM p-wave annihilation mediated by $X$ should be minor during DM freeze-out, e.g., with the $X$'s contribution $\lesssim$ 10\%. In this case, the DM-electron scattering  cross section becomes less than/or similar to 10\% of the scalar DM-electron scattering case obtained by Ref. \cite{Jia:2016uxs}, i.e., a more broad range of the cross section allowed. We look forward to the MeV scale DM search via the future DM-electron scattering experiment.

\acknowledgments \vspace*{-3ex}

%\begin{acknowledgements}

This work was supported by the National Natural Science Foundation of China under Contract No. 11505144, and the Longshan academic talent research supporting program of SWUST under Contract No. 17LZX323.

%\end{acknowledgements}

\end{document}